%% file: main.tex
\documentclass[reprint,amsmath, amssymb, prx, longbibliography, nofootinbib, floatfix]{revtex4-2}
\usepackage[colorlinks]{hyperref} 
\usepackage{float}
\makeatletter
\let\newfloat\newfloat@ltx
\makeatother
\usepackage{algorithm}
\usepackage{algpseudocode}

\usepackage[colorlinks]{hyperref}   
\usepackage{qcircuit}

\smallskip  
\def\<{{\langle }}
\def\>{{\rangle }}

\def\ket#1{|#1\rangle}
\def\bra#1{\langle#1|}

\smallskip

\usepackage{color}
\usepackage{graphicx}
\usepackage{dcolumn}
\usepackage{bm}

\usepackage{amsmath}

\begin{document} 
\title{Re-QGAN: an optimized adversarial quantum circuit learning framework}

\thanks{This manuscript has been authored by UT-Battelle, LLC, under Contract No. DE-AC0500OR22725 with the U.S. Department of Energy. The United States Government retains and the publisher, by accepting the article for publication, acknowledges that the United States Government retains a non-exclusive, paid-up, irrevocable, world-wide license to publish or reproduce the published form of this manuscript, or allow others to do so, for the United States Government purposes. The Department of Energy will provide public access to these results of federally sponsored research in accordance with the DOE Public Access Plan.}%

\author{Sandra Nguemto}
\affiliation{Department of Mathematics, University of Tennessee, Knoxville, TN 37996, USA}

\author{Vicente Leyton-Ortega} 
\email[Corresponding author: ]{leytonorteva@ornl.gov}
 \affiliation{Computational Sciences and Engineering Division, Oak Ridge National Laboratory, Oak Ridge, TN 37831, USA}

\date{\today}
\begin{abstract}
Adversarial learning represents a powerful technique for generating data statistics. Its successful implementation in quantum computational platforms is not straightforward due to limitations in connectivity, quantum operation fidelity, and limited access to the quantum processor for statistically relevant results. Constraining the number of quantum operations and providing a design with a low compilation cost, we propose a quantum generative adversarial network design that uses real Hilbert spaces as the framework for the generative model and a novel strategy to encode classical information into the quantum framework. We consider quantum generator and discriminator architectures based on a variational quantum circuit. We encode classical information by the stereographic projection, which allows us to use the entire classical domain without normalization procedures. For low-depth ans\"atze designs, we consider the real Hilbert space as the working space for the quantum adversarial game. This architecture improves state-of-the-art quantum generative adversarial performance while maintaining a shallow-depth quantum circuit and a reduced parameter set. We tested our design in a low resource regime, generating handwritten digits with the MNIST as the reference dataset. We could generate undetected data (digits) with just 15 epochs working in the real Hilbert space of 2, 3, and 4 qubits. Our design uses native quantum operations established in superconducting-based quantum processors and is compatible with ion-trapped-based architectures.
\end{abstract}

                              
\maketitle    

\input{intro} 
\input{pipeline}

\input{applications}

\input{structure}
\input{discussion}

\section*{Data availability}
All data needed to evaluate the conclusions are available from the corresponding author upon request.

\acknowledgements  
This work was performed at Oak Ridge National Laboratory, operated by UT-Battelle, LLC under contract DE-AC05-00OR22725 for the US Department of Energy (DOE). Support for the work came from the DOE Advanced Scientific Computing Research (ASCR) Accelerated Research in Quantum Computing (ARQC) Program under field work proposal ERKJ354. 

S.N. was supported in part by the U.S. Department of Energy, Office of Science, Office of Workforce Development for Teachers and Scientists (WDTS) under the Science Graduate Laboratory Internship program.

\section*{Author contributions}
S.N. and V.L.-O designed the qGAN workflow and the adversarial algorithm. V.L.-O. designed the stereographic encoding. S.N. and V.L.-O wrote the code for the analysis and figures, analyzed the experimental results, and contributed to the final version of the manuscript. 

\section*{Competing interests}
The authors declare that there are no competing interests. 
\appendix
\input{2qubitPrep}

\end{document}

%% file: intro.tex
\section*{}

Quantum Machine Learning \cite{biamonteQuantumMachineLearning2017,schuldMachineLearningQuantum2019} has been concerned with applying concepts and tools from quantum computation and quantum information to enhance the treatment and processing of classical machine learning tasks. Several approaches demonstrate a level of advantage when the quantum processor (QPU) is used in classical-quantum hybrid architectures, from the implementation of generative tasks using the QPU as a parameterized sample generator \cite{benedetti2019a} until classification procedures exploiting an entangled measurement scheme in the QPU to enhance the discrimination efficiency \cite{google2022}. One case is the adversarial networks framework upgrading, known as quantum generative adversarial neural networks (qGANs). This quantum framework uses the QPU to build discriminative and generative models based on parameterized quantum circuits (PQC), where the data is encoded into a quantum state and processed by quantum operations. The adversarial game structure (quantum or classical) requires a classical database known as the real-data, a generative model (generator) that can produce data known as fake-data, and a discriminative model (discriminator) that can measure the difference between real and fake. The generator and discriminator compete; from one side, the generator tends to produce fake-data that can be treated as real-data by the discriminator. On the other side, the discriminator improves the way of distinguishing real- from fake-data.

qGANs were introduced in 2018 in two companion papers \cite{dallaire-demersQuantumGenerativeAdversarial2018,lloydQuantumGenerativeAdversarial2018}, presenting a roadmap to build their architecture and sketch their operational structure and advantages over their classical counterpart. Since then, several works have proposed different qGANs architectures in several paradigms and for other applications. In addition, these works have tackled the various problems linked to implementing and training qGANs. For instance, quantum conditional GANs \cite{liuHybridQuantumclassicalConditional2021} that outperformed their classical counterpart, quantum style-based GAN (style-qGAN) \cite{bravo-prietoStylebasedQuantumGenerative2021}, where the quantum generator model is optimized to generate Monte-Carlo events for high-energy physics applications, quantum dissipative GAN \cite{beerDissipativeQuantumGenerative2021} based on dissipative neural networks, entanglement qGAN (EQ-GAN) \cite{niuEntanglingQuantumGenerative2021} that entangles target and generated states to achieve the Nash equilibrium in the adversarial game, and qGANs to model classical continuous distributions \cite{romeroVariationalQuantumGenerators2021}. 

Other works treat implementation matters, like strategies using shallow quantum circuits and post-measurement procedures to discriminate the generator's output  \cite{steinQuGANGenerativeAdversarial2021}, studies on the convergence of qGANs and score functions to enhance the convergence time \cite{bracciaHowEnhanceQuantum2021}, and encoding protocols to efficiently load random distributions into $n$-qubit quantum states \cite{zoufalQuantumGenerativeAdversarial2019}. Most of the proposed qGANs schemes require a non-trivial compilation for their implementation into commercial QPUs, and a resetting according to quantum hardware limitations in connectivity and quantum circuits executions. 

One crucial aspect not present in the classical framework is the classical information encoding into quantum information. The adversarial game works in unison with the encoding procedure. Some efforts on this matter cover the effects of data encoding on the expressivity of quantum models \cite{schuldEffectDataEncoding2021}, and the impact of the encoding on binary quantum classifiers \cite{larose2020robust}, where they evaluate decision boundaries for different encoding strategies under noise influence. A typical procedure is to shrink the classical information into the Hilbert space and return the processed quantum information to the classical world. If the encoding is not one-to-one, we can incur loss and tergiversation. In this work, we address the question of how to design simple qGAN that suit the quantum hardware limitation and at the same time leverage its capabilities, and how we can encode classical information in a quantum state without incurring transformations that yield imprecise information in the de-encoding process.    

We contribute to this rich body of work by proposing a novel classical-quantum encoding based on the stereographic projection and a real quantum state qGAN (re-qGAN) that, in general, requires an optimal value of parameters, according to the size of the classical data, and only single-qubit rotations and CXs to be achieved. This design produces less computationally complex quantum neural networks compared to previous architectures. The real amplitude quantum state preparation, in turn, improves training time and feasibility on near-term quantum devices.

%% file: pipeline.tex
\section{Results} \label{sec:pipeline}

\subsection*{Quantum Adversarial Game}
We propose a generative adversarial quantum circuit learning for an unsupervised learning task. In this work, we test the strategy in image generation using a supervised loss for an unsupervised learning task. We consider the standard structure for quantum generative adversarial algorithms, a real-data agent that encodes the sample of classical information (training sample) into quantum information, a generator that prepares quantum states that eventually will be decoded into classical data, and a discriminator that measures how qualitatively different is the generator’s output is from the training dataset. We describe the input data and present the quantum adversarial model in the following.

\begin{algorithm}
\caption{Quantum adversarial game}\label{algo:gan}
\begin{algorithmic}
    \State {\bf Input}: \begin{itemize}
        \item $\cal C$: dataset $\{\vec{u}_0, \cdots, \vec{u}_N \}$, with $\vec{u}_k = (u_1, \cdots, u_M) $ 
        \item EPOCHS: Number of learning cycles  
    \end{itemize}    
    \State {\bf Initialization:}  
    \begin{itemize}
        \item SI encoding ${\cal C} \rightarrow {\cal C}_Q$ (Eq. \ref{eq:SIencoding})  
        \item $n = \lceil \log_2 (M) \rceil$, number of qubits 
        \item set $\vec{\theta_D} \in [-\pi,\pi)^{2^{n+1} -1}$ arbitrarily
        \item set $\{ \vec{\theta}_G^{\, k}  \in [-\pi, \pi)^{2^{n} -1}$ \} arbitrarily,  $k = 1,\cdots, N$
        \item build ${\cal D}^n$ and ${\cal G}^n$ (Fig. \ref{fig:workflow}(c) and (d))
    \end{itemize}
    \For{$i =1$ to EPOCHS} (Fig. \ref{fig:workflow}(b))
    \State ${\cal D}^n$ training: $\vec{\theta}_D \gets {\rm arg\, min}_{\theta_D} {\cal L}_{D}$ (Eq. \ref{eq:ld})
    \State ${\cal G}^n$ training: $\{\vec{\theta}_G^{\, k} \} \gets {\rm arg\, min}_{ \{\vec{\theta}_G^{\, k} \}  } {\cal L}_{G}$ (Eq. \ref{eq:lg})
    \EndFor
    \begin{itemize}
        \item prepare ${\cal Q}_Q = \{ |q_0\rangle, \cdots , |q_N\rangle \}$ from $\{\vec{\theta}_G^{\, k} \}$ (Eq. \ref{eq:generator})
        \item SI de-encoding ${\cal Q}_Q \rightarrow {\cal Q}$ (Eq. \ref{eq:de-encoding})
    \end{itemize}
    \State {\bf Output}: \begin{itemize}
        \item ${\cal Q}$: artificial dataset $\{\vec{q}_0, \cdots, \vec{q}_N \}$  
    \end{itemize}
\end{algorithmic}
\end{algorithm}

We consider as input data a collection of images ${\cal C} = \lbrace \vec{u}_1, \cdots, \vec{u}_N \rbrace$, that follow an equivalence relation. For simplicity, we convert the images into grayscale. Thus, the input data can be easily formatted into arrays. The vectorized images are compressed by a principal component analysis (PCA) to be projected into the $n$-qubit real Hilbert (a hypersphere) by the inverse stereographic projection. The stereographic projection allows the use of the projected information, from the real plane where the input data lies to the hypersphere, to define the set of quantum state ${\cal C}_Q = \lbrace \ket{u_1}, \cdots, \ket{u_N} \rbrace$ that will represent the classical data. With the state amplitude representation of the classical data, the discriminator's goal is to differentiate this set from any output produced by the generator.

The generator and discriminator agents are designed using parameterized quantum circuits ${\cal G}(\mathbf{\theta})$ and ${\cal D}(\mathbf{\theta})$, respectively, that prepare arbitrary quantum states with real amplitudes. The generator follows a special design that minimizes the number of parameters to the optimal point $2^n - 1$ for full expressivity,  given $n$ the number of qubits. On the other hand, the discriminator entangles any pair of quantum states with real amplitudes and uses an ancillary qubit whose measurement decides whether an input state is equivalent to the training sample. 

The discriminator is calibrated with $C_{Q}$ tuning the parameters $\vec{\theta}_D$ in ${\cal D}$ such that the quantity
\begin{equation} \label{eq:sigma}
    \sigma(\ket{s},\vec{\theta}) = \bra{\psi_D^{s}(\vec{\theta})} I \cdots I \cdot Z \ket{\psi_D^{s}(\vec{\theta})} 
\end{equation}
becomes $1$ for each $\ket{\psi_D^{s}(\vec{\theta})} = {\cal D}(\vec{\theta}) \ket{s} \otimes \ket{0}_{ancilla}$ with $\ket{s} \in {\cal C}_Q$, and -1 for every other state $\ket{s} \not \in {\cal C}_Q$. In concurrency, the generator is updated to prepare quantum states more equivalent to the sampling training set with the goal of being undetected by the discriminator, this procedure defines a set of parameters $\theta_D$. We prepare a sample of fake-data produced by ${\cal G}$ defining the set of parameters $\lbrace \vec{\theta}^{\, 1}_G, \cdots,  \vec{\theta}^{\, N}_G\rbrace$; for a short notation $\{\vec{\theta}^{\, k}_G\}_N $. As a first step, the fake sample is compared with the training dataset to calibrate the discriminator. To capture the generator-discriminator adversarial dynamics, we define the cost functions for the discriminator and generator training as 
\begin{eqnarray}
  {\cal L}_D(\vec{\theta}_D;\{\vec{\theta}^{\, k}_G\}_N) &=& \frac{1}{2N} \sum_{l = 1}^N \left( \left[ 1 - \sigma (\ket{u_l},\vec{\theta}_D)\right]^2 \right. \label{eq:ld} \\ \nonumber && +  \left. \left[1+\sigma( \ket{\psi_G (\vec{\theta}_G^{\, l})},\vec{\theta}_D ) \right]^2 \right),  \\  
    {\cal L}_G(\{\vec{\theta}^{\, k}_G\}_N; \vec{\theta_D}) &=& \frac{1}{N} \sum_{l = 1}^N 
    \left[ 1 - \sigma ( \ket{\psi_G ( \vec{\theta}_G^{\, l})},\vec{\theta}_D ) \right]^2 \label{eq:lg}, \nonumber \\ 
\end{eqnarray}
respectively, with $\ket{\psi_G (\vec{\theta}_l)} = {\cal G}(\vec{\theta}_l) \ket{0\cdots 0}$ as the $l$th state prepared by the generator. The minimization of the discriminator cost function, ${\cal L}_D$, sets a configuration where the elements of ${\cal C}_Q$ and the artificial data from ${\cal G}$ are recognized with $\sigma = 1$ and $\sigma = -1$, respectively. The minimization of ${\cal L}_G$ gives the parameter configuration for the generator ${\cal G}$ that produces $\sigma = 1$. In every epoch cycle, the discriminator is optimized to detect the generator's output, and the generator is optimized to fool the discriminator. Once a stop criterion is reached, either a maximum number of iterations or convergence, the set $\{ \vec{\theta}^{\, k}_G \}_N$ will generate new data indistinguishable to ${\cal C}$ after a SI de-encoding procedure. In Algorithm \ref{algo:gan} we present the pseudo-code for the quantum adversarial game. This procedure considers the compressed information when the PCA is applied as to the input information. In the next section, we shall provide more details about the Discriminator's and Generator's training.

%% file: applications.tex
\subsection*{Generating MNIST handwritten digits}

\begin{widetext}
\begin{minipage}{\linewidth}
\begin{figure}[H]
\centering
\includegraphics[width=0.9 \linewidth]{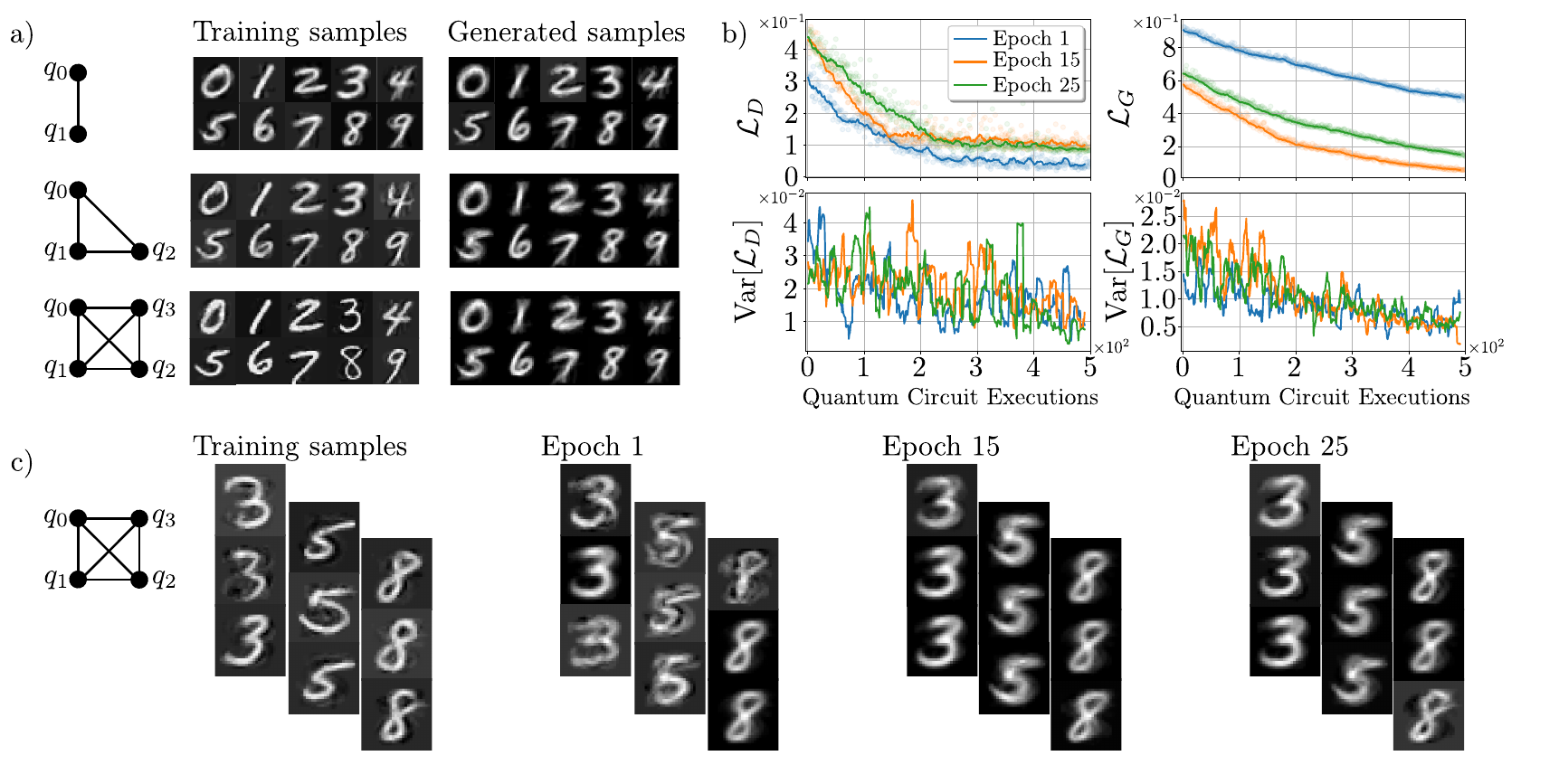}
\caption{MNIST benchmark generation problem. In this Figure, we present the re-qGAN generated digits using a subset of the MNIST as the reference dataset. In a) we present the generated digits for each qubit configuration, we compress the reference dataset according to the number of qubits; see Table \ref{tab:mnist_config} for the level of compression $n_{PCA}$ and re-qGANs characteristics used. In panel b), the learning for the discriminator and generator at various epochs (Epoch 1, 15, and 25), for four qubits, learning the digit `8'. Each epoch represents one learning cycle, with 500 CMA-ES iterations. In the top row of this panel, we have the evolution concerning the number of CMA-ES iterations of the discriminator and generator's loss functions, $\mathcal{L}_D$ and $\mathcal{L}_G$, respectively. In the bottom row, we have the evolution of the variances of $\mathcal{L}_D$ and $\mathcal{L}_G$ throughout the optimization process. In panel c), we present representative elements of the generated samples by re-qGAN for the digits `3', `5' and `8' using four qubits and 50 training samples per digit at different epochs, namely, epochs 1, 15, and 25. Here, the cumulative explained variance for the PCA transformation is $\sim 75 \%$.\label{fig:digits}}
\end{figure}
\end{minipage}
\end{widetext}

Generating handwritten digits using the MNIST dataset constitutes a vast common recognition task and a vital benchmark problem in machine learning. In this case, we considered a subset of the MNIST as the training dataset for our re-qGAN model. We compress the MNIST subset to fit the amount of information we can encode for a given number of qubits, so we test the performance of the qGAN in generating low-resolution images. In this experiment, we consider 25 epochs and a subset of 20 images from the MNIST to generate a new digit. In all the numerical experiments, we consider a reduced number of qubits (from 2 to 4) and low resolution ($28\times 28$) black and white images in the training dataset compressed by the principal component analysis (PCA) at a degree of explained variance that depends on the number of qubits used, i.e., for $n$ qubits the number of components $n_{\rm PCA}$ must be less than or equal to $2^n-1$. On the other hand, the maximal number of components depends on the training dataset size. In Table \ref{tab:mnist_config}, we show relevant details of the numerical experiments. In Figure \ref{fig:digits}(a), we present the numerical experiment results and generated images for different sizes in the quantum representation.  

In this experiment, the re-qGAN could generate undetectably (from the discriminator viewpoint) samples at the 15th epoch cycle (top row in Fig. \ref{fig:digits}(b)), where we observe a good trade-off between the optimal values of ${\cal L}_D$ and ${\cal L}_G$. In Figure \ref{fig:digits}(b) we show the variance of ${\cal L}_D$ and ${\cal L}_G$, and according to the CMA-ES search mechanism, the generator output is closer to a converged value that the discriminator, a favorable scenario in the adversarial game. The fewer epochs for good quality outputs represent an improvement compared to the results in recent proposals, which generally require around 100 epochs to generate undetectable samples \cite{steinQuGANGenerativeAdversarial2021}. From this observation, we can infer the ability of re-qGAN to learn from a reference database and its potential implementation on limited access quantum devices. 

In addition, where we consider a subset of the MNIST as a reference dataset, the generated digit's resolution and quality do not scale with the number of qubits (Fig. \ref{fig:digits}(a) and (c)), which is related to the size of the subset chosen for the learning task. In this case, we found an attractive characteristic to explore, how the subset size works in unison with the number of qubits to generate undetectable data.  Looking at the samples generated by re-qGAN for four qubits (Fig. \ref{fig:digits}(c)), we observe a diverse selection of outputs, which indicates that the qGAN is effectively learning the original distribution instead of simply producing random outcomes. We also deal with lower resolution training samples than in Figure \ref{fig:digits}(a) when increasing the training dataset size because of the decreased cumulative explained variance (CEV) in the PCA process. Better resolution outputs would likely require finding a good trade-off between sample size and CEV.
\begin{table}[ht]
    \centering
    \begin{tabular}{c|cc|cc | cc}
    &  \multicolumn{2}{c|}{${\cal G}^n$} & \multicolumn{2}{c|}{${\cal D}^n$} & \multicolumn{2}{c}{PCA} \\
    $n$  & $Y_\theta$s & CXs & $Y_\theta$s & CXs & $n_{\rm PCA}$ & CEV \\ \hline
     2  & 3   & 1  & 9  & 4  & 3  & 51 \% \\
     3  & 7   & 4  & 18 & 11 & 7  & 76 \% \\
     4  & 15  & 11 & 35 & 26 & 15 & 96 \% 
    \end{tabular}
    \caption{List of parameters used in the numerical experiments. We present the number of parameters ($Y_\theta$s) and entangling gates (CXs) used in every case. In addition, we present the number of components used and the cumulative explained variance (CEV). The training dataset is a subset of 20 images from MNIST.}
    \label{tab:mnist_config}
\end{table}

%% file: structure.tex
\section{methods}\label{sec:structure}

\subsection{Classical dataset encoding}

A standard classical-quantum encoding procedure requires mapping an element of the classical data as quantum state amplitude, i.e., we can associate $\vec{u} \in {\cal C}$ with an element $\vec{v} = (v_1(\vec{u}), \cdots , v_N(\vec{u}))$, whose components define the amplitudes of a quantum state, therefore $||\vec{v}|| \stackrel{!}{=} 1$. For instance, in the standard amplitude encoding $v_j = u_j/||\vec{u}||$, every single element in the set $\{ \alpha \vec{u}: \alpha \in \mathbb{R} \}$ will have the same representation $\vec{v}$, losing information and creating a non one-to-one map between the classical and quantum information.  We employed the inverse of the stereographic projection (IS map) to dodge this.

The IS map associates vectors $\vec{u}$ in the real plane $\mathbb{R}^{2^n - 1}$ with vectors $\vec{v}$ in $\mathbb{S}^{(2^n - 1)} - \{ {\rm N} \}$ (see Fig. \ref{fig:workflow}(b)), using the relationship 
\begin{eqnarray}\label{eq:SIencoding}
    v_i &=& \frac{2 u_i}{||\vec{u}||^2 + 1}, \ {\rm for} \ i = 1, \cdots, 2^n-1  \nonumber \\
    v_{2^n} &=& \frac{||\vec{u}||^2 - 1}{||\vec{u}||^2 + 1}. 
\end{eqnarray}
With this, we can define a unique quantum state 
\begin{equation*}
    |v\rangle = v_0 \ket{0 \cdots 00} + v_1 \ket{0\cdots 0 1} + \cdots + v_{2^n - 1} |1\cdots 1 1\rangle \ .
\end{equation*}

 The stereographic encoding can accurately represent any vector in the real plane, meaning that encoding is one-to-one. Nevertheless, any variation in the real plane scales with the inverse $||\vec{u}||^{-2}$, i.e., the area is not conserved; we can see this from the relation: $dA_{\rm sphere} \propto dA_{\rm plane} / (1+ ||\vec{u}||^2)$ with $dA_{\rm sphere}$ and $dA_{\rm plane}$ are the area elements of the sphere and the plane, respectively. This feature could make it hard to distinguish the quantum state representation of large vectors, $||\vec{u}|| >> 1$. Without loss of generative power, we scale the training dataset to be close, in magnitude, to a unitary value, keeping $||\vec{u} || \sim 1$, where $dA_{\rm sphere} \sim dA_{\rm plane}$ \cite{dubrovin1992modern}. In the IS map, we need to establish one scaling factor for all elements in the classical dataset.
Once we process the information in the Hilbert space, the answer can be de-encoded back to the real plane by 
\begin{equation}\label{eq:de-encoding}
    u_i = v_i/(1 - v_{2^{n}}), \ {\rm for} \ i = 1, \cdots , 2^{n} -1 \ .
\end{equation}

\begin{widetext}
\begin{minipage}{\linewidth}
\begin{figure}[H]
\centering
    \includegraphics[width=0.7\textwidth]{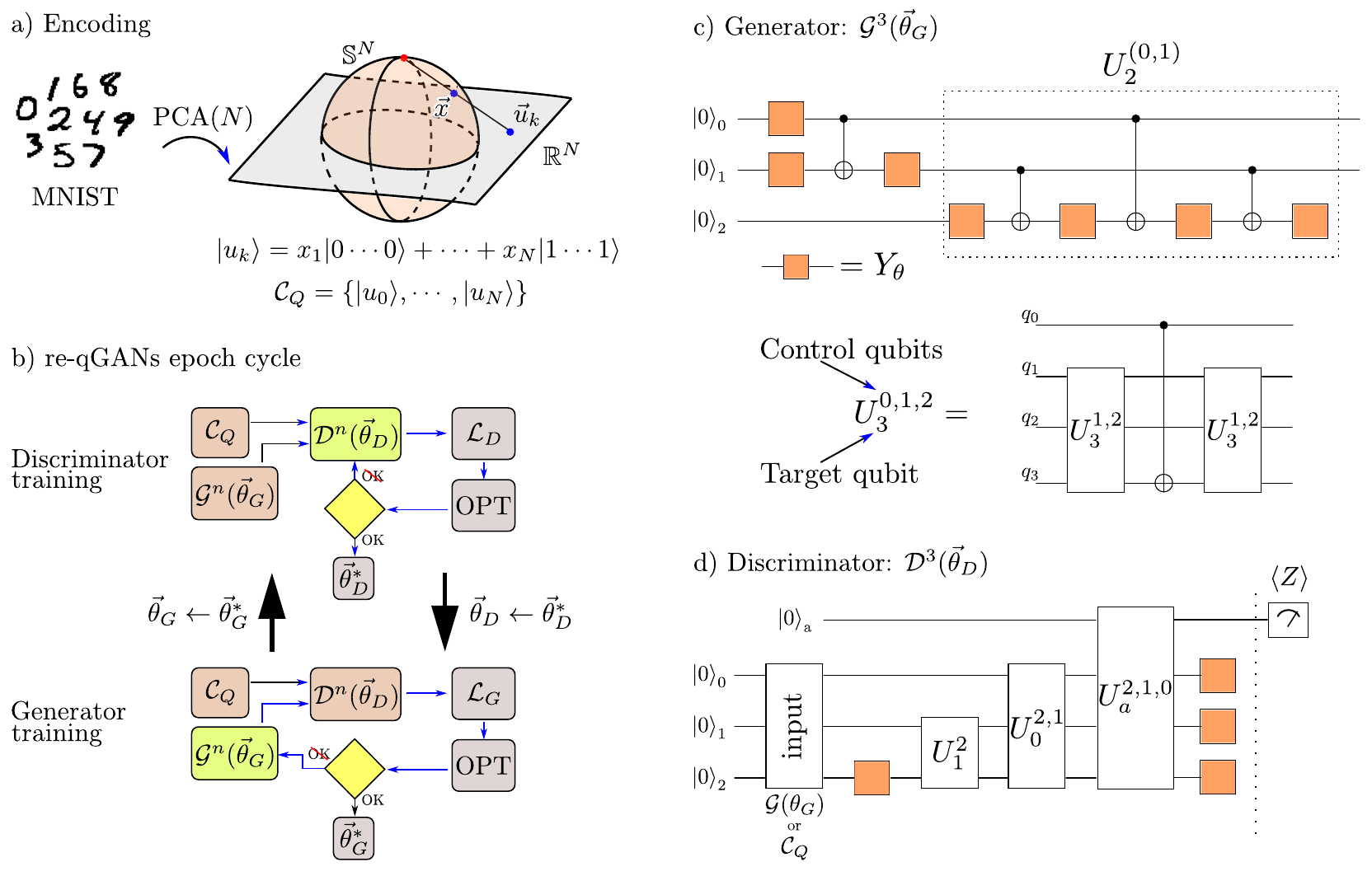}
    \caption{Sketch of the quantum adversarial game elements. In a), we present an illustration of the SI encoding of a subset of the MNIST. We compress the images using the principal component analysis (PCA) and project them onto a hypersphere to define the representative quantum state amplitudes. On-hardware implementation requires the preparation of every element of the set ${\cal C}_Q$ by using the PQC used for the generator. b) Once we get the real-data set, ${\cal C}_Q$, we proceed to the sequential discriminator and generator training. Inside, OPT is the classical optimization algorithm that updates the discriminator or generator parameters. The stop criteria depend on the optimizer choice. In c) we present the generator quantum circuit, the ansatz prepares a 2-qubit state and entangles it with the third qubit through a uniformly controlled rotation (UCR). This PQC can be extended to more qubits by applying a cascade of UCRs. For the 4-qubit case, we need to add the UCR $U_3$ growing the PQC parameter set in $2^3$ more elements. At the bottom, the relation between $U_3^{0,1,2}$ and $U_3^{1,2}$. The recurrence can be used for higher levels of UCRs (see text for more details). d) Three-qubit discriminator ansatz, ${\cal G}^3$. The ansatz transforms the generated quantum state or ${\cal C}_Q$ to a quantum state in a higher dimensional Hilbert space. The measurement in the ancillary qubit serves as a metric to establish the nature of the input. \label{fig:workflow}}
\end{figure}
\end{minipage}
\end{widetext}

\subsection{Quantum state generator and discriminator}
We have established the quantum representation of the classical dataset; now, we shall proceed to introduce a parameterized quantum circuit (PQC) able to prepare the elements in $C_Q$ and new quantum states that the discriminator will test.

For the quantum state generator and discriminator, we consider the real amplitude version of the PQC introduced by M\"ott\"onen \emph{et al.} \cite{mottonenTransformationQuantumStates2004, mottonenDecompositionsGeneralQuantum2005}. The generator is designed to prepare an $n$-qubit arbitrary quantum state with the optimal number of parameters, $2^n - 1$. Figure \ref{fig:workflow}(b) presents the 3-qubit PQC based on $Y_\theta$ rotations and CX gates. The first part of the ansatz prepares an arbitrary two-qubit quantum state. This design was studied in \cite{perdomoEntanglementTypesTwoqubit2021} as an optimal preparation. The second part introduces a two-fold uniformly controlled rotation (UCR) $U_2^{0,1}$. In general, the $m$-fold UCR  $U_{q_m}^{q_0,\cdots, q_{m-1}}(\theta_1,\cdots,\theta_{2^m})$ plays a vital role the state preparation. This operator applies the rotation $Y_{\theta_l}$ on the target qubit state $\ket{q_m}$ when $\ket{q_0 \cdots q_{m-1}} = \ket{l^*}$ with $l^*$ as the $m$-bit representation of $l$, for $l = 1, \cdots, 2^m$. The $m$-fold UCR is built recursively following 
\begin{eqnarray}
     U_{q_m}^{q_0,\cdots, q_{m-1}}(\theta_1,\cdots,\theta_{2^m}) 
     && = \nonumber \\  && U_{q_m}^{q_1,\cdots, q_{m-1}}(\theta_1,\cdots,\theta_{2^{m-1}}) \nonumber \\  &\cdot& CX_{q_0,q_m} \nonumber \\ &\cdot& U_{q_m}^{q_1,\cdots, q_{m-1}}(\theta_{2^{m-1}+1},\cdots,\theta_{2^{m}}), \nonumber \\
\end{eqnarray}
for $m > 1$, where $U_{q_1}^{q_0}(\theta_1,\theta_2) = [IY]_{\theta_1} CX_{q_0q_1} [IY]_{\theta_2}$. The bottom in Figure \ref{fig:workflow}(c) shows how to generate $U_3^{0,1,2}$ from $U_2^{0,1}$. The $n$-qubit generator is built with the application of UCRs at different levels, 
\begin{eqnarray}\label{eq:generator}
 {\cal G}^n(\theta_1, \cdots, \theta_{2^{n}-1}) &=& [YI\cdots I]_{\theta_1} \nonumber \\
 &\cdot& U_{q_1}^{q_0}(\theta_2,\theta_3) \nonumber \\
 &\cdot& U_{q_2}^{q_0,q_1}(\theta_4,\cdots, \theta_7) \nonumber \\
 &\vdots& \nonumber \\
 &\cdot& U_{q_{n-1}}^{q_0,\cdots,q_{n-2}}(\theta_{2^{n}-2^{n-2}},\cdots,\theta_{2^{n}-1}). \nonumber \\
\end{eqnarray}
The number of parameters scale as $2^n -1$ and the number of $CX$ gates as $2^n - n -1$. The relationship between the quantum state amplitudes generated by ${\cal G}^n$ with the parameters $\vec{\theta}_G$ defines a parameterization of $\mathbb{S}^{2^n -1}$. For more details, see Appendix \ref{app:qstateprep}.

\begin{figure}[ht]
    \centering
    \includegraphics[width=\linewidth]{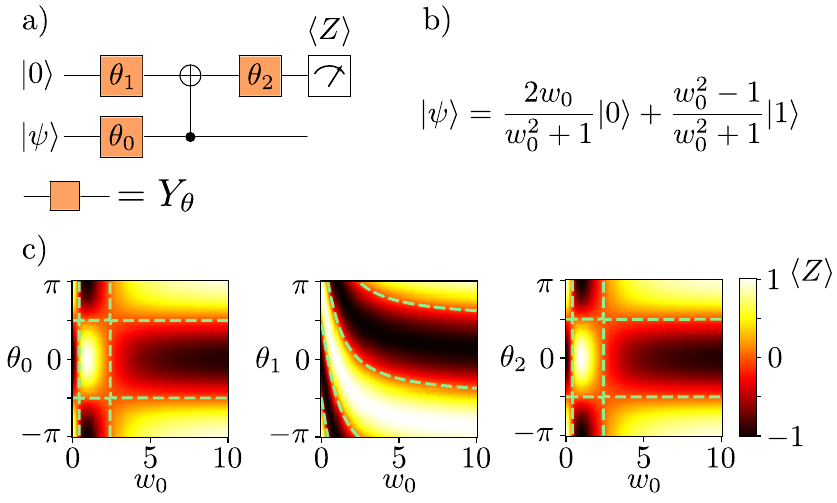}
    \caption{The expressive power of the 2-qubit discriminator. We represent the discrimination power by its projection on the characteristic plane for each discriminator parameter. In a) the quantum circuit for the two-qubit discriminator, b) the SI encoding model for a single-qubit, and c) heat plots showing how the discriminator can separate data classes, identified by the values of $\langle Z \rangle = \pm 1$.}
    \label{fig:disc_boundaries}
\end{figure}

For the discriminator's ${\cal D}(\theta_D)$ structure, we consider an ancillary qubit whose measurement decides whether an input state belongs or not to the quantum dataset ${\cal C}_Q$. In Figure \ref{fig:workflow}(d), we present the four-qubit PQC design, three qubits for the input to discriminate and a fourth qubit to measure. The discriminator aims to project an arbitrary quantum state in $\mathbb{S}^{(2^n - 1)}$ into a Hilbert space $\mathbb{S}^{(2^{n+1} - 1)}$. For that, we consider the same structure used for the generator with extra local rotations in the last layer. This makes the number of parameters scale as $2^{n+1} + n-1$ and the number of CX gates as $2^{n+1} - n - 2$. To illustrate the discrimination mechanism behind ${\cal D}^n$, consider the single-qubit quantum state discrimination. We consider a two-qubit discriminator and classical data domain in the real line, from 0 to 10 (see Figure \ref{fig:disc_boundaries}). Following the SI encoding, see Fig. \ref{fig:disc_boundaries}(b), we show in Fig. \ref{fig:disc_boundaries}(c) the different decision boundaries the discriminator defines when one of its parameters is active, i.e., $\neq 0$. The decision boundary is defined by $\langle Z \rangle = 0$ (see dotted green lines in Fig. \ref{fig:disc_boundaries}(c)).

\subsection{Interaction and optimization}

The discriminator's interaction with ${\cal C}_Q$ and the generator's output is as follows. Initially, we calibrate the discriminator by minimizing ${\cal L}_D (\vec{\theta}_D; \{\vec{\theta}^{\, k}_G\}_N)$ with respect to $\vec{\theta}_D$. Once we calibrate the discriminator, we use the optimal parameter values $\theta_D^*$ in the generator optimization, i.e. $\vec{\theta}_D \leftarrow \vec{\theta}_D^*$. In this step,  the algorithm minimizes ${\cal L}_G (\{\vec{\theta}^{\, k}_G\}_N;\vec{\theta}_D^*)$, obtaining the optimal configuration $\{\vec{\theta}^{\, k}_G\}^*_N$ that will be used to update $\{\vec{\theta}^{\, k}_G\}_N$ in the next epoch, i.e. $\{\vec{\theta}^{\, k}_G\}_N \leftarrow \{\vec{\theta}^{\, k}_G\}^*_N$. In both training processes, discriminator and generator, the algorithm iterates for a fixed number of steps or quantum circuit executions or until the learning reach a local minimum and the costs, ${\cal L}_G$ and ${\cal L}_D$, do not decrease.  

We consider the Covariance Matrix Adaptation Evolution Strategy (CMA-ES), a gradient-free optimization algorithm, to update the generator and discriminator parameters in each learning process. This optimization algorithm uses the principle of biological evolution, where the repeated interplay of variation and selection yields a global minimum. The evolution strategy starts with an initial population of $\Theta = \{\vec{\theta}_1, \cdots \vec{\theta}_P\}$, with $P = 4 + \lfloor 2 \log M \rfloor$ and $\vec{\theta}_k = (\theta_1, \cdots , \theta_M)$. In our case, $M$ is the number of PQC parameters. In each generation (execution of $P$ quantum circuits), a new population $\Theta'$ is generated by variation (according to a multivariate normal distribution) of the current parental population $\Theta$. Then, some individuals in $\Theta'$ are selected to become the next generation parents based on their ${\cal L}_{D/G}$ value. Over the generation sequence, sets of parameters with better and better ${\cal L}_{D/G}$ values are generated. If the population has enough information to estimate the covariance matrix reliably, the optimizer updates the covariance matrix that helps compute the new mean value for the distribution presented in $\Theta'$. For a deeper discussion about stochastic research algorithms, see Ref. \cite{hansen2014}. In the numerical experiments, we use the CMA-ES python implementation \cite{hansen2019pycma}. 

%% file: discussion.tex
\section{Discussion}\label{sec:discussion}

We addressed the question of building a quantum adversarial model per the limitations of commercial quantum processors. We proposed re-qGAN, a re-imagined and optimized architecture for quantum generative adversarial networks, using low-budget resources regarding quantum circuit depth and executions. To avoid any possible loss of information in the classical encoding, we proposed a novel strategy based on stereographic projection. The SI encoding offers a one-to-one map to encode classical data into a real Hilbert space. We can extend this encoding to complex values by embedding the $n$-qubit Bloch sphere in a $\mathbb{R}^{2^{n+1}-1}$, defining a $\mathbb{S}^{2^{n+1}-1}$, and associating the points of this hyper-sphere with points in the real plane $\mathbb{R}^{2^{n+1}-2}$. In doing so, we increase the size of the real-data dimension we can encode per qubit and simultaneously the number of CXs and local operations in the quantum state preparation. To keep the depth of the quantum circuit at a low level and to avoid an unnecessary boost of quantum imperfection in an eventual on-hardware implementation, we stick with the real amplitude constraint even though it needs more qubits to encode the classical information.

Additionally, as illustrated in figure \ref{fig:disc_boundaries}, the more affluent input class enabled by the stereographic encoding positively affects the discriminator's classification ability. However, since the inverse stereographic projection is a conformal map, i.e., it does not preserve distances, fluctuations around the north pole can be amplified in the real plane bringing error in the de-encoding process. We can consider a re-scaling factor for the classical input to keep the projection onto the hyper-sphere away from the north pole. 

In the quantum adversarial game, The generator is designed for real amplitude quantum state preparation. We require fewer parameters but more qubits than we would if we considered complex amplitude quantum states to encode classical information. This approach is motivated by the limited set of native gates available in physical devices and, on the other hand, to minimize the noise effects related to the quantum circuits' depth. The relationship between the generator's parameters and the quantum amplitudes defines a hyper-sphere parameterization as illustrated in the 2 and 3 qubit cases in Appendix \ref{app:qstateprep}. By constructing a generator that covers the entire real Hilbert space, we get a good trade-off between expressive power and computational resource requirements for the adversarial model. 

Agents trained in an adversarial game, both in classical or quantum machine learning \cite{chengGenerativeAdversarialNetworks2020}, do not always converge to Nash equilibrium with gradient descent \cite{mazumdarGradientBasedLearningContinuous2020}. Therefore, by building an architecture using fewer parameters, we can update the generator and discriminator by gradient-free optimization procedures without error propagation in consecutive evaluations present in gradient-based optimizations. The classical optimization performance depends on the choice of the loss function \cite{leyton2021robust}. In this case, we consider a least-square-based loss function (Eqs. \ref{eq:ld} and \ref{eq:lg}), with $\pm 1$ as target values. The configuration of the loss functions pushes the discriminator to converge to a local minimum, under adversarial training, that yields the desired outcome of not being able to classify generated samples as fake data.

In the discriminator's design, we considered the projection of the input quantum state into a higher dimensional Hilbert space. In Figure \ref{fig:disc_boundaries}, we observe an excellent nonlinear behavior in the way the discriminator can separate data classes, where the mean value $\langle Z \rangle$ determines the data class in the ancillary qubit.  

We tested our qGAN architecture in the learning task of generating images of handwritten digits, using a subset of the MNIST dataset as our reference database. We successfully generated handwritten digit images using 2,3 and 4 qubits. We use a genetic algorithm (CMA-ES) to update the generator's and discriminator's parameters due to the reduced number of parameters in our model. the re-qGAN algorithm convergences after a fewer number of epochs, in comparison to state-of-the-art qGAN models, such as \cite{steinQuGANGenerativeAdversarial2021}, while using fewer resources in general. Its ability to learn distributions in a few epochs, using a reduced number of parameters and a small training data set, points to its feasibility on real quantum hardware, with potential advantages such as quantum error robustness, from using fewer operations and qubits. Additionally, the re-qGAN structure uses native quantum operations in most of the superconducting quantum platforms \cite{ibmq}, with a small compilation effort in trapped ions platforms \cite{ionq}. This further points to its feasibility on commercial devices.

Counter-intuitively, the re-qGAN's output resolution does not improve with the number of qubits. We can explore this feature by studying the non-trivial correlation between the size of the reference dataset and the number of qubits. Furthermore, even though re-qGAN's architecture theoretically provides flexibility to add more qubits and thus train it on more complex data, scaling it on quantum hardware will require a good trade-off between computational resources and algorithmic efficiency. More importantly, re-qGAN can be easily implemented on a real quantum machine, such as trapped-ion-based quantum processors, to test its real-world efficiency and benchmark the hardware capabilities.

%% file: 2qubitPrep.tex
\section{preparation of an arbitrary two-qubit quantum state with real amplitudes} \label{app:qstateprep}

To clarify the relationship between the parameters $\vec{\theta}$ and the real amplitudes $x_1, ...., x_{2^n}$, we consider the two and three-qubit cases. The two-qubit case is straightforward and was studied in \cite{perdomoEntanglementTypesTwoqubit2021}. There, the amplitudes follow 
\begin{eqnarray}
 x_1 &=& c_{\theta_1} c_{\theta'_1} , \ x_2 = c_{\theta_1} s_{\theta'_1} \nonumber \\
 x_3 &=& s_{\theta_1} c_{\theta'_2} , \ x_4 = s_{\theta_1} s_{\theta'_2}
 \label{eq:twoqbs}
\end{eqnarray}
with $H_2(\theta_2, \theta_3)^T = (\theta'_1, \theta'_2)$. Where, $H_{k}$ stands for the Hadamard matrix of order $k$, $c_\theta = \cos (\theta/2)$, and $s_\theta = \sin (\theta/2)$. The transformation \eqref{eq:twoqbs} defines a parameterization for $\mathbb{S}^3$, for more details see Ref. \cite{perdomoEntanglementTypesTwoqubit2021}. For the three-qubit case, the amplitudes follow a pattern

\begin{eqnarray}\label{eq:threeqbs}
 x_1 &=& c_{\theta_1} c_{\theta'_1} c_{\theta''_1}, \ 
 x_2 = c_{\theta_1} c_{\theta'_1} s_{\theta''_1}, \nonumber \\
 x_3 &=& c_{\theta_1} s_{\theta'_1} c_{\theta''_4}, \
 x_4 = c_{\theta_1} s_{\theta'_1} s_{\theta''_4},\nonumber \\
 x_5 &=& s_{\theta_1} s_{\theta'_2} s_{\theta''_2},\
 x_6 = s_{\theta_1} s_{\theta'_2} c_{\theta''_2},\nonumber \\
 x_7 &=& s_{\theta_1} c_{\theta'_2} s_{\theta''_3},\
 x_8 = s_{\theta_1} c_{\theta'_2} c_{\theta''_3}, 
\end{eqnarray}
where, $H_2(\theta_2, \theta_3)^T = (\theta'_1, \theta'_2)$, and $H_4(\theta_4, \cdots ,\theta_7)^T = (\theta''_1, \cdots, \theta''_4)^T$. Defining a parameterization for ${\mathbb{S}^7}$, therefore, for three (two) qubits ${\cal G}^{3(2)}$ covers the entire real Hilbert space. By induction, we can establish that ${\cal G}^{n}$ can prepare an arbitrary quantum state. 

An arbitrary real amplitude quantum state $\ket{\psi} = x_1 \ket{00} + x_2 \ket{01} + x_3 \ket{10} + x_4 \ket{11}$ in the 2-qubit Hilbert space is prepared following \ref{eq:twoqbs}. This map between the parameter space and the hypersphere is one-to-one. It is also surjective, since we can find for every $\vec{x} \in \mathbb{S}^3$, a vector $\vec{\theta}$ in the parameter space, such that $\vec{\theta} \rightarrow \vec{x}$ \cite{perdomoEntanglementTypesTwoqubit2021}. Since this is true for $\mathbb{S}^3$, an induction argument will show that this is the case for the n-qubit Hilbert space.